\documentclass[useAMS,usenatbib]{mn2e}

\usepackage{aas_macros,graphicx,multirow,rotate,amssymb}

\def\src{GRB\,140206A}
\def\int {\emph{INTEGRAL}}
\def\swi {\emph{Swift}} 
\def\fermi {\emph{Fermi}}

\title[GRB 140206A]{GRB 140206A: the most distant polarized Gamma-Ray Burst}
\author[D. G\"{o}tz et al.]{D. G\"{o}tz$^{1}$\thanks{E-mail:
diego.gotz@cea.fr}, P. Laurent$^{2}$, S. Antier$^{1}$, S. Covino$^{3}$, P. D'Avanzo$^{3}$, V. D'Elia$^{4,5}$, \newauthor A. Melandri$^{3}$
\smallskip\\
$^{1}$AIM (UMR 7158 CEA/DSM-CNRS-Universit\'e Paris Diderot) Irfu/Service d'Astrophysique, Saclay, F-91191 Gif-sur-Yvette Cedex, France\\
$^{2}$APC (UMR 7164 CEA/DSM/Irfu, Universit\'e Paris Diderot, CNRS/IN2P3, Observatoire de Paris) 10, rue Alice Domon et L\'eonie Duquet,\\ 75205 Paris Cedex 13, France\\
$^{3}$INAF -- Osservatorio Astronomico di Brera,  Via E. Bianchi 46, 23807 Merate (LC), Italy\\
$^{4}$INAF-Osservatorio Astronomico di Roma, Via Frascati 33, I-00040 Monteporzio Catone, Italy\\
$^{5}$ASI-Science Data Center, Via del Politecnico snc, I-00133 Rome, Italy
}
\begin{document}

\date{Accepted . Received ; in original form }

\pagerange{\pageref{firstpage}--\pageref{lastpage}} \pubyear{2013}

\maketitle

\label{firstpage}

\begin{abstract}

The nature of the prompt $\gamma$-ray emission of Gamma-Ray Bursts (GRBs) is still far from being completely elucidated. The measure of linear polarization is a powerful tool that can be used to put further constraints on the 
content and magnetization of the GRB relativistic outflows, as well as on the radiation processes at work.

To date only a handful of polarization measurements are available for the prompt emission of GRBs.
Here we present the analysis of the prompt emission of \src, obtained with \int/IBIS, \swi/BAT, and \fermi/GBM. Using \int/IBIS as a Compton polarimeter we were able to constrain the linear polarization level of the second peak of this GRB as being larger than 28\% at 90\% c.l.

We also present the GRB afterglow optical spectroscopy obtained at the Telescopio Nazionale Galileo (TNG), which allowed us the measure the distance of this GRB, z=2.739. This distance value together with the polarization measure obtained with IBIS, allowed us to derive the deepest and most reliable limit to date ($\xi <$1$\times$10$^{-16}$) on the possibility of Lorentz Invariance Violation, measured through the vacuum birefringence effect on a cosmological source.

\end{abstract}


\section{Introduction}

Gamma-Ray Bursts (GRBs) are transient sources whose duration spans from ms up to thousands of seconds in some cases. Most of their energy is emitted in the $\gamma$-ray band around a few hundreds of keV and they appear unpredictably at random directions on the whole sky, making their understanding challenging. In fact, despite the recent progresses in the GRB field, mainly obtained thanks to GRB-dedicated instrumentation, like the one on the \textit{Swift} and \textit{Fermi} satellites \citep[see e.g.][]{gehrels09,zhang14}, the nature of the prompt emission of GRBs is still not completely clear. On the other hand, much information could be obtained from the long-lived GRB afterglows in the X-ray, optical, and radio bands. GRBs have been proven to be of cosmological origin, with their redshifts, $z$, distributed in the range [$0.1,\sim9$], and several of them are now firmly associated with Supernovae of type Ib/c, and hence with the collapse of massive stars.

GRBs emit during a few seconds a huge amount of isotropic equivalent energy, $E_{\rm iso}$, that spans from 10$^{50}$ to 10$^{54}$ erg \citep[e.g.][]{amati08}, making them the most luminous events in the Universe, temporarily outshining all other sources. However, GRBs are likely collimated sources and the true emitted energy is then reduced to about 10$^{51}$ erg \citep{frail01,bloom03,ghirlanda12}. Nonetheless, the exact geometry and content of this collimated jet, as well as its magnetization are not elucidated yet, and the details of the mechanism leading to the $\gamma$-ray emission are still not completely clear. Models include
unmagnetized fireballs, where the observed emission could be
produced by relativistic ($\Gamma \ga 100$) electrons accelerated in
internal shocks propagating within the outflow \citep{rees94}, and span to pure
electromagnetic outflows where the radiated energy comes from magnetic
dissipation \citep{lyutikov06}. Intermediate cases with mildly
magnetized outflows are also envisaged \citep[e.g.][]{spruit01}.
%Even in the case of an unmagnetized fireball, a local magnetic field
%in the emission region, generated by the shocks, is necessary if the
%dominant process is synchrotron radiation from relativistic electrons.

Polarization measurements could add an additional constraint with respect to spectral and timing information, and indeed in the recent years, some measurements of polarization during the prompt emission of GRBs in the hundreds of keV energy range have been attempted using \int/IBIS,
\int/SPI, and IKAROS/GAP
\citep[][]{kalemci07,mcglynn07,mcglynn09,gotz09,gotz13,yonetoku11,yonetoku12}. 
%These
%measurements have the potential to provide new inputs about the strength and scale of magnetic fields in the relativistic jets, as well as about the radiative mechanisms at work during the GRBs. 
Thanks to these measurements the open questions mentioned above could be tackled. In fact, even if globally incoherent, in the case where
the magnetic field is mainly transverse and locally highly ordered,
i.e. has a local coherence scale which is larger than the typical size
$\sim R / \Gamma$ of the visible part of the emitting region, a synchrotron polarized 
signal can still be detected. 
This scenario has been favoured in the case of GRB 041219A \citep{gotz09}, 
where a time resolved analysis could be performed, and the rapid polarization angle variations
could be explained by the variation of the bulk Lorentz factors $\Gamma$ of the emission regions. On the other hand, for GRBs for which just a time integrated measure is available, different scenari like
the case of a random field or an ordered magnetic field parallel to
the expansion velocity, for which the polarization of the detected signal should
vanish, except for the peculiar condition of a jet observed slightly
off-axis \citep[e.g.][]{lazzati04}, cannot be completely excluded.

Further clues on the magnetic structure of GRB jets, but at later times with respect to the prompt emission, came recently thanks 
to the results presented by \citet{mundell13}. They report the detection of 
a high level of linear polarization (28$\pm$4\%) in the early optical afterglow of GRB 120308A, indicating the presence of large scale magnetic field surviving long
after the initial explosion. In that case the emission has been modelled as due mainly to the reverse shock taking place when the relativistic ejecta interact with the GRB ambient medium. Indeed the GRB early afterglow emission is produced by a combination of the radiation of the forward and reverse shock, and the reverse shock tests the magnetic structure of the inner part of the jet, just like prompt emission, but at slightly later times, when the prompt emission produced internally to the jet is over.

Finally we note that polarization measures 
in cosmological sources are also a valuable tool for fundamental physics experiments:
Lorentz Invariance Violation (LIV) arising from the
phenomenon of vacuum birefringence can be constrained as recently shown by \citet{fan07},
\citet{laurent11a}, \citet{stecker11}, \citet{toma12}, and \citet{gotz13}.
 
Here we present the prompt emission analysis of \src\ obtained with
\int, \swi, and \fermi/GBM, as well as its polarization measurements
obtained with \int\ (section \ref{sec:analysis}). We also present the
spectroscopy of the GRB afterglow obtained with the
Telescopio Nazionale Galileo (TNG) (section \ref{sec:TNG}) and discuss
our results, including the LIV limits (section \ref{sec:liv}) we can obtain from this GRB, in section \ref{sec:discussion}.

\section{Data Analysis and Results}
\label{sec:analysis}

\src\ has been detected by the \int\ Burst Alert System
\citep[IBAS;][]{ibas} on February 2$^{nd}$ 2014, and localized to
R.A. = $09^{h}41^{m}13.03^{s}$ Dec.=
$+66^{\circ}45^{\prime}54.7^{\prime\prime}$, with an 90\%
c.l. uncertainty of 0.8$^{\prime}$ \citep{gotz14}. The burst has been
also been detected and localized by \swi\
%\ \footnote{The \int\ GCN
 % notice was issued 20 seconds before the \swi\ one.} 
\citep{lien14} and the GBM on board \fermi\ \citep{vonkienlin14}. A bright optical
afterglow at a position consistent with the prompt one, peaking at
about the 15$^{th}$ magnitude was reported by several telescopes
\citep{oksanen14, oates14, yurkov14, xu14, volnova14, sonbas14,
  davanzo14, masi14, saito14, kopac14, quadri14, toy14}. The
brightness of the optical counterpart allowed to measure the redshift
of the GRB ($z \sim$ 2.7) independently by two groups
\citep{malesani14, delia14}. In the following sections we present the
analysis of the prompt $\gamma$-ray emission of \src\ and of its
optical afterglow.

\subsection{IBIS/ISGRI}

IBIS \citep{ibis} is a coded mask telescope on board the \int\
satellite \citep{integral}. It is made by two superposed pixellated detector
layers, ISGRI \citep{isgri} working in the 15 keV--1 MeV energy range,
and PICsIT \citep{picsit}, working in the 200 keV -- 10 MeV energy
range. Here we restrict our analysis to the ISGRI detector
plane. Indeed, due to satellite telemetry limitations, PICsIT
spectral--imaging data are temporally binned over the entire duration
of an \int\ pointing (typically 30–-45 minutes) and hence they are not
suited for studies of GRBs, while for PICsIT spectral--timing data no
proper response matrix is available yet.

\src\ has been detected by IBIS/ISGRI at the very beginning of the
\int\ orbit, while still close to the radiation belts. So, due to the
high count rate induced by the residual particle flux, not all the
data could be transmitted to the ground, especially while the GRB was
at its peak. This is why we decided to include in this paper also the
\swi\ and \fermi/GBM data analysis in order to have a complete picture
of the GRB. Otherwise e.g. the GRB peak flux and fluence would have
been underestimated.

Using the \int\ Off-line Scientific Analysis (OSA) software v. 10.0 we
extracted the ISGRI light curve of \src\ in 3 s time bins, which is
the shortest time bin for which sufficient data are available. As can
be seen from Fig. \ref{fig:lc}, most of the time bins are empty due to
the telemetry loss. Nevertheless due to the high flux of the GRB the
time bins for which an analysis is possible have rather high
statistics reaching up to 2500 counts/bin. This allowed us to extract
two spectra, corresponding to the two main peaks of the GRB, namely
from 07:17:20.0 to 07:17:50.0 U.T. for the first peak and from
07:18:10.0 to 07:18:40 U.T. for the second peak. These spectra have
been used for the common spectral fit with the other instruments, see
below. 

We note that telemetry bandwidth limitations do not affect the polarimetric
results (see below), since Compton events packets are prioritized in the IBIS
telemetry transmission.

\subsection{Swift/BAT} 
 
The \swi/BAT \citep{swift, bat} data have been downloaded through the \swi\ public archive\footnote{http://swift.gsfc.nasa.gov}, and analysed with the tools provided by HEASARC v. 6.15.1, and the latest version of CALDB. BAT standard products, including the light curve shown in Fig. \ref{fig:lc}, have been extracted using {\tt batgrbproduct}. Previously, the mask weightening which produces background-subtracted lights curves and spectrum has been validated using {\tt fkeyprint}. The detector quality map of the two peaks has been computed taking into account the same time intervals as for ISGRI, with the help of {\tt batgbin} which creates a detector plane image, and {\tt batdetmask}, retrieving the appropriate detector quality map from CALDB. The two spectra of the two main peaks, have been derived using {\tt batbinevt} and several corrections needed to fit the two spectrum with {\tt xspec} (see Section \ref{sec:spanalysis}) have been applied using {\tt batupdatephakw} and {\tt batphasyserr}. The appropriate response matrices have been derived using {\tt batdrmgen}. The BAT data have been used to measure the GRB $T_{90}$ duration resulting to be 93.2$\pm$13.5 s in the 15--300 keV energy band. 

\subsection{GBM}

\fermi/GBM \citep{gbm} data have been obtained through the \fermi\
SSC\footnote{http://fermi.gsfc.nasa.gov/ssc/}, and analysed using the
RMFIT v. 432
package\footnote{http://fermi.gsfc.nasa.gov/ssc/data/analysis/rmfit/}. Using
the quick look data, we chose the NaI and BGO detectors for which the
GRB signal is stronger. This corresponds to the NaI detectors number
8 and 11, and to the BGO number 1. The light curves of the three
detectors have been computed by subtracting the background fitted
using a fourth degree polynomial function over time intervals before
and after the GRB, excluding the GRB itself.

\begin{figure}
\centering
\includegraphics[angle=0,width=8.5cm]{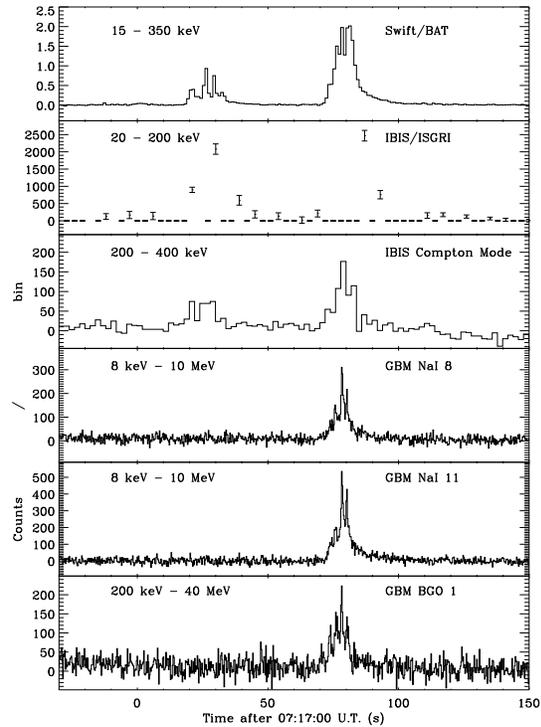}
\caption{Light curves of \src\ as seen by different detectors. The
  detector name and energy range are reported in each panel. Note that
  except for BAT for which the counts are reported per detector pixel,
  the global count rate for each instrument is shown. The time bin is
  1 s for BAT, 3 s for IBIS/ISGRI, 2 s for IBIS Compton Mode and 0.25 s
  for the GBM detectors. The first peak of the GRB is missing in the GRB data due to Earth occultation.}
      \label{fig:lc}
\end{figure}

As can be seen from Fig. \ref{fig:lc}, only the second peak of the GRB
has been detected in the GBM data, because of the occultation of the
source by the Earth during the first peak. A spectrum of the second
peak has been extracted for the three detectors mentioned above, using
again the same time interval as for ISGRI. The spectra have been
exported to PHA format, and the latest available response matrices
have been obtained through the CALDB database.

\subsection{Joint Spectral Analysis}

\label{sec:spanalysis}

BAT provides the most complete data set for \src. That is why we used
these data to derive the GRB peak flux and fluence. The GRB peak flux
in the 15--350 keV energy band measured over 1 s is 20.7$\pm$1.0 ph
cm$^{-2}$. The GRB fluence measured over the entire burst duration is
about 2$\times$10$^{-5}$ erg cm$^{-2}$. The average BAT spectrum is
well fitted ($\chi^{2}$/d.o.f.=64.7/72) by a power law with an
exponential high-energy cut-off, with a photon index
$\Gamma$=1.1$\pm$0.15 and a cut-off energy $E_{c}$=114$^{+47}_{-26}$
keV. The errors are reported at 90\% c.l.

\begin{figure*}
%\centering
%\hspace{-1.2cm}
\includegraphics[width=8.5cm]{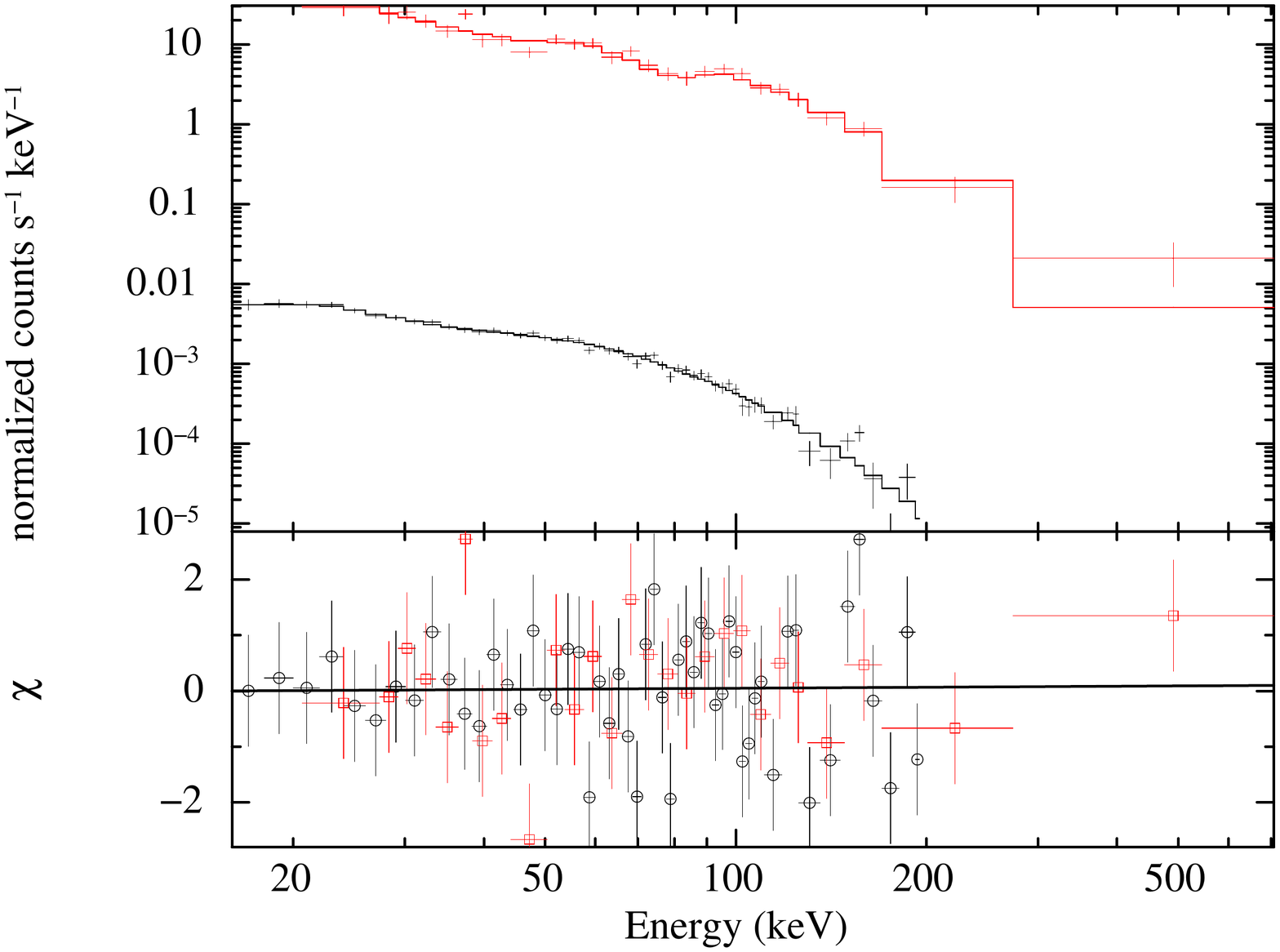}
\includegraphics[width=8.5cm]{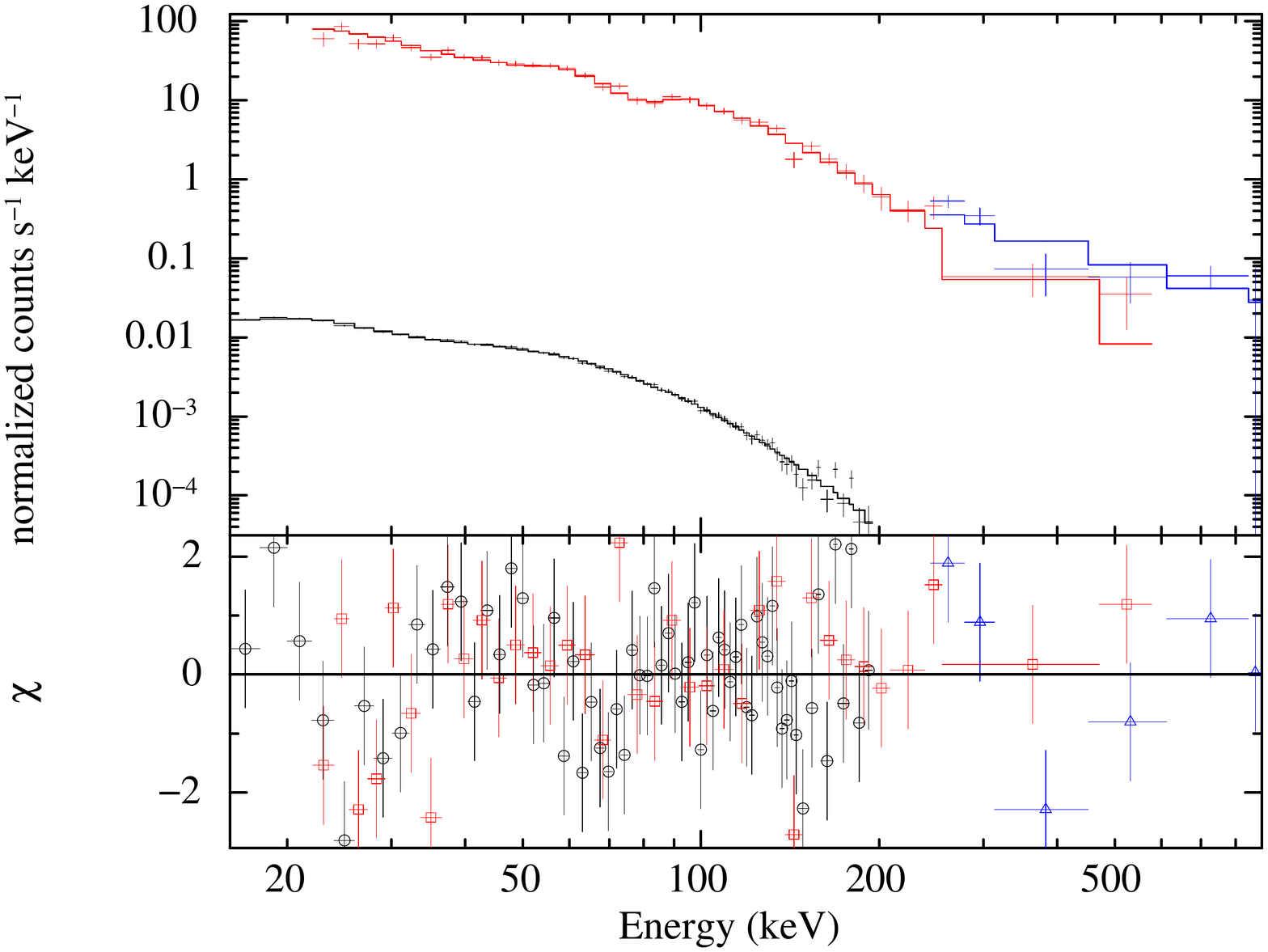}
\caption{Left: Joint Swift/BAT (black) and IBIS/ISGRI (red) spectrum of the first peak of \src. The upper panel shows the recorded counts and the best fit
  model, while the lower panel shows the residuals with respect to the
  model using circles for BAT and squares for ISGRI. Right: Joint Swift/BAT (black), IBIS/ISGRI (red), GBM/BGO (blue)
  spectrum of the second peak of \src. The upper panel shows the recorded counts and the
  best fit model, while the lower panel shows the residuals with
  respect to the model using circles for BAT, squares for ISGRI, and
  triangles for the BGO.}
      \label{fig:secondsp}
\end{figure*}

The spectra of the two peaks derived for the different instruments
(see above), have been fitted simultaneously using {\tt xspec}
v. 12.7.0 \citep{xspec}. For the first peak only ISGRI and BAT spectra
have been used. A constant multiplicative factor has added in order to
account for cross-calibration uncertainties and ISGRI data loss. For
the joint fit of the first peak, see Fig. \ref{fig:secondsp}, a simple
power law could be excluded ($\chi^{2}$/d.o.f.=133.3/109), and a
cut-off power law represented a better model for the data
($\chi^{2}$/d.o.f.=112/108). A fit using a Band function \citep{band93}
did not increase further the quality of the fit
($\chi^{2}$/d.o.f.=112/107).

For the second peak we used BAT, ISGRI, and GBM BGO data. GBM NaI data
have been excluded due to their lower statistical quality with respect
to ISGRI and BAT. In this case, thanks to the BGO data extending the
spectral coverage to higher energies, the best fit model is
represented by a Band model ($\chi^{2}$/d.o.f.=148/121), since the fits using a single
power law ($\chi^{2}$/d.o.f.=465/123) or a cut-off power law ($\chi^{2}$/d.o.f.=162/122) turn out
to be less adapted to the data.  
The spectral
fitting results for both peaks are reported in Table
\ref{tab:spec}. One can see that the GRB peak energy, $E_{p}$,
decreases with time, and that both values are on the soft end of the
peak energy distribution of the GRBs observed with \int, \fermi/GBM or
BATSE \citep{bosnjak13}.

\begin{table}
\centering
\caption{Spectral fitting results for \src. Errors are given at 90\% c.l.} 
\label{tab:spec}
\begin{tabular}{cccc}
\hline
\textit{First Peak}\\
Photon Index $\Gamma$ & Cut-off energy $E_{c}$  & -- &$E_{p}$\\ 
\\
1.13$\pm$0.16 & 150$_{-44}^{+92}$ keV & -- & 130$_{-38}^{+80}$ keV \\ 
\hline
\textit{Second Peak}\\
$\alpha$ & $E_{0}$ & $\beta$ & $E_{p}$\\ 
\\
0.94$\pm$0.08 & 92$\pm$16 keV & 2.0$_{-0.3}^{+0.2}$ & 98$\pm$17 keV \\
\hline
\end{tabular}
\end{table}

\subsection{Polarization}

%IBIS \citep{ibis} is a coded mask telescope on board the \int\ satellite \citep{integral}. It is made by two pixellated detector layers, ISGRI \citep{isgri} working in the 15 keV--1 MeV energy range, and PICsIT \citep{picsit}, working in the 200 keV -- 10 MeV energy range. 
The two superposed pixellated detector layers permit to IBIS to be
used as a Compton telescope by measuring the properties of the photons
(time, energy and position) interacting in both planes.  Thanks to the
polarization dependency of the differential cross section for Compton
scattering, linearly polarized photons scatter preferentially
perpendicularly to the incident polarization vector. Hence a Compton
telescope can be used also as a polarimeter, and IBIS allowed us to date to
detect polarization in five different bright objects, the Crab nebula
\citep{forot08}, the black hole binary Cyg X--1 \citep{laurent11b},
GRB 041219A \citep{gotz09}, GRB 061122 \citep{gotz13}, and GRB 120711A
(Martin-Carrillo et al., in prep.). In this work we adopt the same
analysis technique as described in these references.

Due to the nature of Compton scattering, one can expect an azimuthal
distribution of the scattered photons on the telescope lower plane of
the form

\begin{equation}
N(\phi)=S[1+a_{0}\cos 2(\phi-\phi_{0})],
\label{eq:azimuth}
\end{equation}

where $S$ is the average source flux, $a_{0}$ is the flux amplitude modulation, $\phi$ the azimuthal scattering angle, $P.A. = \phi_{0} - \pi /2 + n \pi$ is the polarization angle (where 0$^{\circ}$ corresponds to the North and 90$^{\circ}$ to the East).
The polarization fraction is defined as $\Pi= a_{0}/a_{100}$, where $a_{100}$ is
the amplitude expected for a 100\% polarized source derived by Monte
Carlo simulations of the instrument \citep[see e.g.][]{forot08}.

%To perform the polarization analysis, we derived the source flux as a
%function of $\phi$, and the scattered photons were then divided in 6
%bins of 30$^{\circ}$. To improve the signal-to-noise ratio in each bin, we took
%advantage of the $\pi$-symmetry of the differential cross section,
%i.e. the first bin contains the photons with
%$0^{\circ}<\phi<30^{\circ}$ and $180^{\circ}<\phi<210^{\circ}$, etc.
%The chance coincidences (i.e. photons interacting in both detectors,
%but not related to a Compton event), have been estimated using the data before the GRB and subtracted from each detector image following the procedure described in
%\citet{forot08}. The derived detector images were then deconvolved to
%obtain sky images, where the flux of the source in each bin is
%measured by fitting the instrumental PSF to the source peak, building
%a so-called polarigram of the source,

\label{sec:results}
%\subsection{Polarization measured with IBIS}
%\label{integral:polar}

%\textbf{***Una volta mi ha detto Costa, quello di SAX, che la principale perplessit\`a riguardo alle misure di polarizzazione di Integral \`e dovuta la fatto che per sorgenti note (Crab?) i risultati sono stati non del tutto soddisfacenti. E' cos\`\i{}? Si pu\`o brevemente menzionare la cosa e perch\`e per noi non \`e un problema?***}

In Fig. \ref{fig:lc} we report the IBIS background-subtracted light
curve of \src, derived using the Compton mode events in the 200--400
keV energy range. Due to the softness of the event, no signal has been
detected above those energies.

We performed the polarization analysis over different time intervals
of the GRB. The best signal-to-noise ratio is obtained over the
07:18:10.0--07:18:30.0 U.T. time interval (corresponding to the second
peak, Compton image SNR=13.5). As can be seen from Fig. \ref{fig:lc}, the GRB first
peak has not enough statistics to perform a sensitive analysis.  In
order to compute $a_{100}$, we used the spectral analysis described
above. Using these spectral parameters, $a_{100}$ has been computed
through Monte Carlo simulations, and turns out to be 0.29$\pm$0.03 (68\% c.l.).

%independent from the energy band, being 0.29$\pm$0.02 in the 250--800 keV energy band, 0.30$\pm$0.03 in the 250--350 keV energy band, and 0.29$\pm$0.03 in the 350--800 keV energy band. 

\begin{figure}
%\centering
\hspace{-1.1cm}
\includegraphics[angle=180,width=10cm]{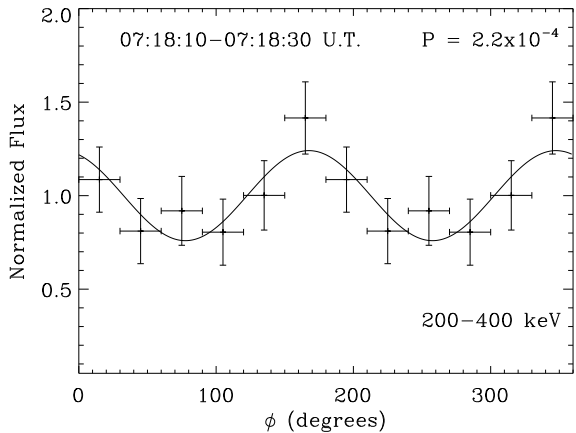}
\caption{Polarigram of \src\ in the 200--400 keV energy band. The crosses
represent the data points (replicated once for clarity)
and the continuous line the fit done on the first 6 points using Eq. \ref{eq:azimuth}. 
The chance probability P of a non-polarized ($<$1\%) signal
is also reported. The normalized flux corresponds to $N(\phi)/S$}.
      \label{fig:polarigram}
\end{figure}

Following the method reported in \citet{gotz13} we first built the source polarigram (i.e. source flux as a function of $\phi$) in the 200--400 keV energy band, see Fig. \ref{fig:polarigram}. We then divided the selected time
interval in smaller energy intervals (200--250 keV; 250--300 keV;
300--400 keV), but only the first energy interval provides a sufficiently
high detection level in order to constrain polarization (SNR=10.6).
We fitted the polarigrams with Eq. \ref{eq:azimuth} using a least
squares technique ($\chi^{2}$/d.o.f.=2.31/2) to derive $a_{0}$ and $\phi_{0}$, see
Fig. \ref{fig:polarigram}. Confidence intervals on $a_{0}$ and
$\phi_{0}$ were, on the other hand, not derived from the fit, since
the two variables are not independent. They were derived from the
probability density distribution of measuring $a$ and $\phi$ from $N$
independent data points over a $\pi$ period, based on Gaussian
distributions for the orthogonal Stokes components (see Eq. 2 in
\citealt{forot08}).

\begin{figure}
%\centering
\hspace{-1.1cm}
\includegraphics[angle=180,width=10cm]{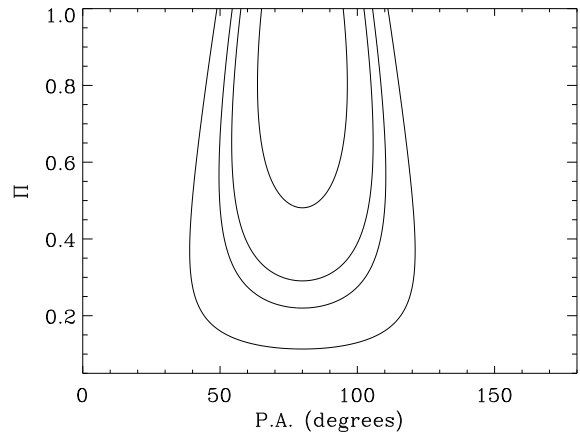}
\caption{The 68\%, 90\%, 95\%, and 99\% (top to bottom) confidence contours for the $\Pi$ and $P.A.$ parameters.}
      \label{fig:errors}
\end{figure}

Over the selected time interval we measure a high polarization level
in the 200--400 keV energy band, deriving a 68\% c.l. lower limit to
the polarization fraction ($\Pi$) of 48\% and the corresponding polarization angle
($P.A.$) is 80$\pm$15$^{\circ}$, see Table \ref{tab:pola}. The 68\%,
90\%, 95\%, and 99\% confidence regions for the two parameters are
shown in Fig. \ref{fig:errors}, where one can see that the 95\%
c.l. lower limit to $\Pi$ is 25\%. For each polarigram we also computed the probability, $P$, the source we measure corresponds to an un-polarized ($\Pi<$1\%) source.
%that the $Pi$ and $P.A.$ values we measure may be produced
%by chance by an un-polarized ($\Pi<$1\%) source. 
This value is reported in 
Fig. \ref{fig:polarigram}.
%So despite the fact that the
%statistics is not high enough to fully constrain the polarization
%fraction, we can exclude that our signal is due to an un-polarized
%source ($\Pi<$1\%) at a probability level $P$ of
%$\sim$2$\times$10$^{-4}$ (also shown in Fig. \ref{fig:polarigram}).
As stated above, the same analysis has been performed in different
energy bands, but only the 200--250 keV energy band shows an
un-polarized probability of 3$\times$10$^{-4}$, while the two others
do not show a statistically significant detection. The derived
parameters for the 200--250 keV energy band are statistically
consistent with the ones of the 200--400 keV energy band.

\begin{table}
\centering
\caption{Polarization measurements of \src.} 
\label{tab:pola}
\begin{tabular}{ccccc}
\hline
Energy band & $\Pi$ (\%)   & P.A. ($^{\circ}$) & $\Pi$ (\%) & P.A. ($^{\circ}$)\\ 
(keV) & (68\% c.l.) & (68\% c.l.) & (90\% c.l.) & (90\% c.l.)\\ 
\hline
200--400 & $>$48 & 80$\pm$15 & $>$28 & 80$\pm$25 \\
%250--350 & $>$65 & 145$\pm$15 & $>$35 & 145$\pm$27 \\
%350--800 &$>$52 & 160$\pm$20 & $>$20 & 160$\pm$38 \\
\hline
\end{tabular}
\end{table}

\section{TNG}
\label{sec:TNG}

The spectroscopy of \src\ was carried out at the Telescopio Nazionale Galileo (TNG) 
using the DOLORES camera in slit
mode, with the LR-B grism \citep{delia14}. This configuration
covers the spectral range $3000-8430$\AA\, with a resolution of
$\lambda/\Delta\lambda = 585$ for a slit width of $1$'' at the central
wavelength $5850$\AA.  The observation started at 2014-02-06T19:53:07,
i.e., $\sim 12.6$ hrs after the GRB, with a total exposure of $1800$
s. The slit position angle was set to the parallatic value.

The spectra were extracted using standard procedures (bias and
background subtraction, flat fielding, wavelength and flux
calibration) under the packages
ESO-MIDAS\footnote{http://www.eso.org/projects/esomidas/} and
IRAF\footnote{http://iraf.noao.edu/}. Ne-Hg and Helium lamps were used for wavelength calibration. A spectrophotometric star could not be acquired the same night of the target, so we used the normalized spectrum for our analysis.
%Ne-Hg or Helium lamps, and
%spectrophotometric stars acquired the same night of the target were
%used for wavelength and flux calibration, respectively.

The TNG spectrum shows several absorption lines that can be interpret
as due to Ly-$\beta$, Ly-$\alpha$,
%{\ion{N}{V}$\lambda\lambda$1238,1242}, {\ion{Si}{II}$\lambda$1260},
%{\ion{O}{I}$\lambda$1302}, {\ion{Si}{II}$\lambda$1304}
%{\ion{C}{II}$\lambda$1334}, {\ion{Si}{IV}$\lambda\lambda$1393,1402},
%{\ion{Si}{II}$\lambda$1526}, {\ion{C}{IV}$\lambda\lambda$1548,1550},
%{\ion{Fe}{II}$\lambda\lambda$1608,1611}, {\ion{Al}{II}$\lambda$1670},
%{\ion{Al}{III}$\lambda\lambda$1854,1862} at a common redshift of
{{N}{V}\,$\lambda\lambda$1238,1242}, {{Si}{II}\,$\lambda$1260},
{{O}{I}\,$\lambda$1302}, {{Si}{II}\,$\lambda$1304},
{{C}{II}\,$\lambda$1334}, {{Si}{IV}\,$\lambda\lambda$1393,1402},
{{Si}{II}\,$\lambda$1526}, {{C}{IV}\,$\lambda\lambda$1548,1550},
{{Fe}{II}\,$\lambda\lambda$1608,1611}, {{Al}{II}\,$\lambda$1670},
{{Al}{III}\,$\lambda\lambda$1854,1862} at a common redshift of
$z=2.739 \pm 0.001$, corresponding to a luminosity distance of 23 Gpc\footnote{Assuming H$_{0}$=71 km/s/Mpc, $\Omega_{M}$=0.27, $\Omega_{\lambda}$=0.73}, and implying an $E_{\rm iso}$ of (2.4$\pm$0.2)$\times$10$^{54}$ erg. In addition, we also detect at the same redshift faint fine structure lines from excited levels of {{Si}{II}*\,$\lambda$1533} and
{{Fe}{II}*\,$\lambda\lambda$1618,1621}. These levels are produced by the GRB light which photoexcites the intervening gas at distances in the range 0.1--2 kpc \citep[see, e.g.][]{prochaska06, vreeswijk07,delia09}. Thus, the excited gas resides in the GRB host, confirming that the GRB is at redshift $z$=2.739$\pm$0.001.

Our redshift determination is in
perfect agreement with the reported value by \citet{malesani14}.

\begin{figure*}
\centering
\includegraphics[angle=-90,width=17cm]{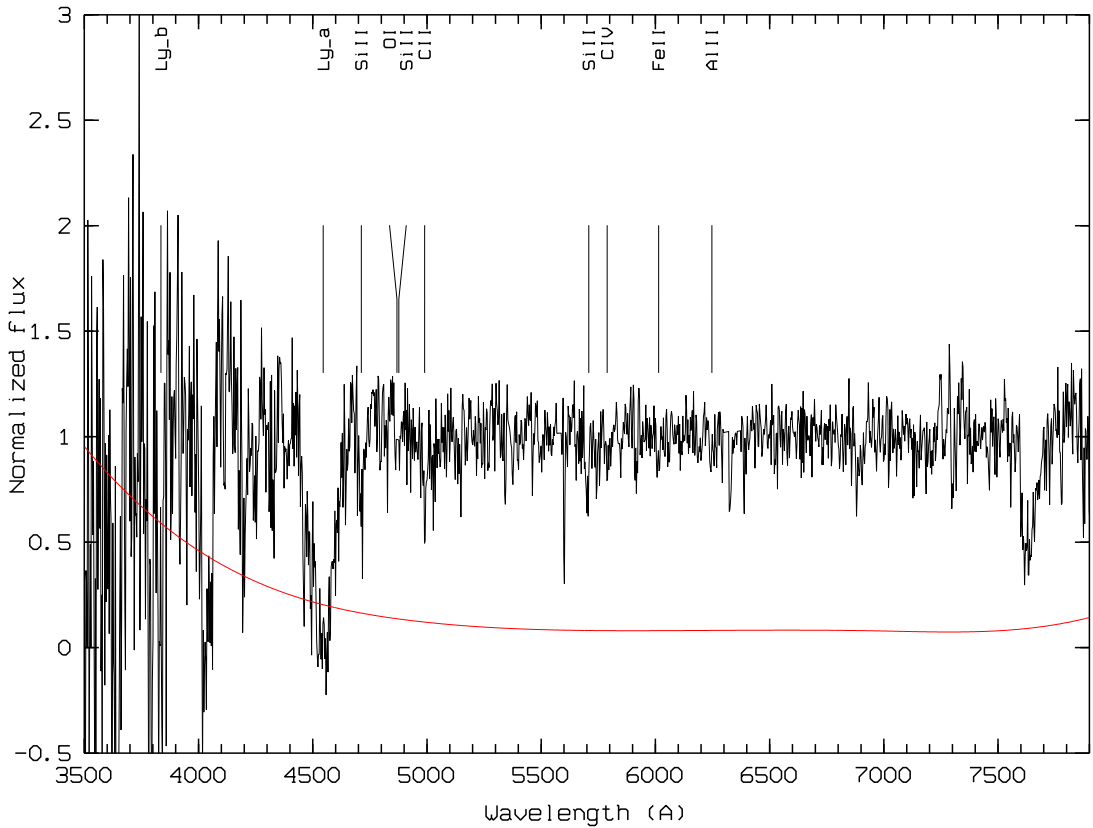}
\caption{The TNG spectrum (black) and its error spectrum (red). Vertical lines mark the strongest absorption features. The Ly-$\alpha$ intervening absorber at z=2.32 is the wide feature at $\sim$ 4000 \AA.}
      \label{fig:TNG}
\end{figure*}

Finally, we also detect a strong intervening Ly-alpha absorber at $z=2.32$. The TNG spectrum with all the absorption features is shown in Fig. \ref{fig:TNG}.

%
%D'Elia et al. 2014, GCN Circ 15802
%
%Malesani et al. 2014, GCN Circ 15800

\section{LIV limits}
\label{sec:liv}

The possible unification at the Planck energy scale of the theory of General Relativity and the quantum theory in the form of the Standard Model requires to quantize gravity, which can lead to fundamental difficulties: one of these is to admit the Lorentz Invariance Violation (LIV)
\citep[e.g.][]{jacobson06,liberati09,mattingly05}

A possible experimental test for such violation is to measure the helicity dependence
of the propagation velocity of photons \citep[see e.g.][and references
therein]{laurent11a}. The light dispersion relation is given in this
case by

\begin{equation}
\omega^{2}=k^{2}\pm\frac{2\xi k^{3}}{M_{Pl}}\equiv\omega^{2}_{\pm}
\label{eq:dispersion1}
\end{equation}

where $E=\hbar\omega$, $p=\hbar k$, $M_{Pl}$ is the Planck Mass, and the sign of the cubic term is determined by the chirality (or circular polarization) of the photons, which leads to a rotation of the polarization during the propagation
of linearly polarized photons. This effect is known as vacuum birefringence.

Equation \ref{eq:dispersion1} can be approximated as follows

\begin{equation}
\omega_{\pm}=\vert k \vert \sqrt{1\pm\frac{2\xi k}{M_{Pl}}}\approx\vert k\vert(1\pm\frac{\xi k }{M_{Pl}})
\label{eq:dispersion2}
\end{equation} 

where $\xi$ gives the order of magnitude of the effect. In practice some quantum-gravity theories \citep[e.g.][]{myers03} predict that the polarization plane of the electromagnetic waves emitted by a distant source rotates by a quantity $\Delta\theta$ while the latter propagates through space, and this as a function of the energy of the photons, see Eq. \ref{eq:rotation}, where $d$ is the distance of the source:

\begin{equation}
\Delta\theta(p)=\frac{\omega_{+}(k)-\omega_{-}(k)}{2}\ d\approx\xi\frac{k^{2}d}{2M_{Pl}}
\label{eq:rotation}
\end{equation}

As a consequence the signal produced by a linearly polarized source, observed in a given energy band could vanish, if the distance is large enough, since the differential rotation acting on the polarization angle as a function of energy would in the end add opposite oriented polarization vectors, and hence in a net un-polarized signal.
But being this effect very tiny, since it is inversely proportional to the Planck Mass ($M_{Pl}\sim$2.4$\times$10$^{18}$ GeV), the observed source needs to be at cosmological distances. The simple fact to detect the polarization signal from a distant source, can put a limit to such a possible violation. This experiment has been performed recently by \citet{laurent11a}, \citet{toma12}, and \citet{gotz13} making use of the prompt emission of GRBs. %Indeed, being GRBs cosmological objects, they are very good candidates to measure and improve those limits.
%, since the latter depend strongly on the frequency and the distance plays a role until $z\sim 1$, but it is less important beyond. 
Indeed, since GRBs are at the same time at cosmological distances, and emitting at high energies, their polarization measurements are highly suited to measure and improve upon these limits.
%\citet{laurent11a}, taking advantage from the polarization measurements obtained with IBIS on GRB 041219A in different energy bands (200--250 keV, 250--325 keV), and from the measure of distance of the source (z$>$0.02 at 90\% c.l., equivalent to a luminosity distance 85 Mpc) were able to set the most stringent limit to date to a possible LIV effect: $\xi <$1.1$\times$10$^{-14}$. We note that, although \citet{toma12} claim to have derived a more stringent limit ($\xi <$8$\times$10$^{-16}$), their measure does not rely on a real measure of the distance of the GRBs they analyse, but they use a distance estimate based on an empirical spectral-luminosity relation \citep{yonetoku10}, whose selection effects, physical interpretation, and absolute calibration are not yet completely understood.

By taking the distance of \src\ we derived above, i.e. 23 Gpc, and if we set $\Delta\theta(k)$ = 90$^{\circ}$ (the fact that we measure the polarization in a given energy band means that the differential rotation should not be grater than this value), we obtain

\begin{equation}
\xi < \frac{2 M_{Pl}\Delta\theta(k)}{(k_{2}^{2}-k_{1}^{2})\ d}\approx 1\times 10^{-16},
\label{eq:xi}
\end{equation}

improving the previous limit \citep{gotz13} by a factor three.

\section{Discussion and Conclusions}
\label{sec:discussion}

We measured the timing and spectral properties of the prompt $\gamma$-ray emission of \src, using \swi/BAT, \fermi/GBM, and \int/IBIS. Using IBIS in Compton mode we were able to measure the linear polarization in the $\gamma$-ray energy band (200--400 keV) during the second and brightest peak of the prompt emission of \src, putting a lower limit on the polarization level of 28\% (90\% c.l.).
%, and exclude an un-polarized signal at a probability level of $\sim$2$\times$10$^{-4}$. 
This measure, follows some recent reports of detections of high (and variable) polarization levels in the prompt emission of a few other GRBs: 041219A by \citet{gotz09,mcglynn07},  061122 by \citet{gotz13, mcglynn09}, 100826A, 110301A and 110721A by \citet{yonetoku11,yonetoku12}, see Table \ref{tab:polasummary}. Although all these measures, taken individually, have not a very high significance ($\gtrsim$3 $\sigma$), they indicate that GRBs are indeed good candidates for highly $\gamma$-ray polarized sources, and that they are prime targets for future polarimetry experiments.
On the other hand, as can be seen from Table \ref{tab:polasummary} the currently available GRB sample does not show extreme spectral characteristics, like e.g. in terms of peak energy, but they are on the upper end of the GRB fluence distribution. This means that, on one hand, this sample may be well representative of the whole GRB population. On the other hand the fluence bias is clearly an instrumental selection effect due to the high photon statistics needed to perform the polarization measurements in IBIS and GAP.

\begin{table*}
\centering
\caption{Summary of recent GRB polarization measurement by IBIS and GAP.} 
\label{tab:polasummary}
\begin{tabular}{cccccc}
\hline
GRB & $\Pi$ (68\% c.l.)  & Peak energy (keV) & Fluence and Energy Range (erg cm$^{-2}$) & $z$ & Instrument\\ 
\hline
041291A & 65$\pm$26\% & 201$^{+80}_{-41}$ & 2.5$\times 10^{-4}$ in 20--200 keV & 0.31$^{+0.54}_{-0.26}$ & IBIS\\ 
06122 & $>$60\% & 188$\pm$17 & 2.0$\times 10^{-5}$ in 20--200 keV & 1.33$^{+0.77}_{-0.76}$ & IBIS\\ 
100826A & 25$\pm$15\%& 606$^{+134}_{-109}$ & 3.0$\times 10^{-4}$ in 20 keV--10 MeV & 0.71--6.84$^{1}$ & GAP\\
110301A & 70$\pm$22\% & 107$\pm$2 & 3.6$\times 10^{-5}$ in 10 keV--1 MeV & 0.21--1.09$^{1}$ & GAP\\
110721 & 84$^{+16}_{-28}$\% & 393$^{+199}_{-104}$ & 3.5 $\times 10^{-4}$ in 10 keV--1 MeV & 0.45--3.12$^{1}$ & GAP\\
140206A & $>$48\% & 98$\pm$17 & 2.0$\times 10^{-5}$ in 15--350 keV & 2.739$\pm$0.001 & IBIS\\ 
\hline
\end{tabular}
\\$^{1}$ redshift based on empirical prompt emission correlations, not on afterglow observations.
\end{table*}

As discussed in \citet{gotz09,gotz13} these polarization features can be explained by synchrotron radiation in an ordered magnetic field \citep{granot03a,granot03b,nakar03}, by the jet structure \citep{lazzati09}, or , independently from the magnetic field structure or the emission processes, by the observer's viewing angle with respect to the jet \citep{lazzati04}, even in the case of thermal radiation from the jet photosphere \citep{lundman14}. In addition the level of magnetization of the jet can also play role \citep{spruit01,lyutikov06}. For instance the ICMART model \citep{icmart}, which implies a magnetically dominated wind launched by the central engine, predicts a decrease of the polarization level during GRB individual pulses, but this hypothesis cannot be tested with the current data. Indeed, as pointed out by \citet{toma09}, the different models are hardly distinguishable relaying only on $\gamma$-ray data, and a result can be achieved only on statistical grounds, i.e. having a sample of several tens of measures at high energies.

On the other hand, the recent detection of a high level ($\Pi$=28$\pm$4\%) of linear optical polarization in the early afterglow of GRB 120308A, allowed \citet{mundell13} to point out the presence of a magnetized reverse shock with an ordered magnetic field, confirming the presence of high magnetic fields in the GRB ejecta, and indicating that
the multi-wavelength approach could be more fruitful while waiting for a dedicated
GRB polarimetry mission, like e.g. POLAR \citep{polar} or POET \citep{poet}. 

Thanks to our TNG spectrum of the GRB afterglow, we were able to precisely measure the distance of our source, $z=2.739$, making of \src\ the most distant GRB for which IBIS was able to measure a polarized signal. Our distance measurement together with the polarization measure obtained with IBIS, allowed us to derive the deepest and most reliable limit ($\xi <$1$\times$10$^{-16}$) to date on the possibility of Lorentz Invariance Violation, measured through the vacuum birefringence effect on a cosmological source. \src\ is namely the first and only GRB for which a polarization measurement of the prompt emission and spectroscopically determined distance are available at once.

\section*{Acknowledgements}

Based on observations with INTEGRAL, an ESA project with instruments and
science data centre funded by ESA member states (especially the PI
countries: Denmark, France, Germany, Italy, Switzerland, Spain), Czech
Republic and Poland, and with the participation of Russia and the USA,
and on observations made with the TNG under programme ID A26 TAC\_63.
ISGRI has been realized and maintained in flight by CEA-Saclay/Irfu with
the support of CNES. This research has made use of data, software and/or web tools obtained from NASA's High Energy Astrophysics Science Archive Research Center (HEASARC), a service of Goddard Space Flight Center and the Smithsonian Astrophysical Observatory.
We acknowledge the financial support of the UnivEarthS Labex program at 
Sorbonne Paris Cit\'e (ANR-10-LABX-0023 and ANR-11-IDEX-0005-02),
of ASI grant I/004/11/0 and of PRIN-MIUR 2009 grants.

\bibliographystyle{mn2e}
\bibliography{biblio}

%\begin{thebibliography}{99}

%\end{thebibliography}

\label{lastpage}

\end{document}